\documentclass[epj,nopacs]{svjour} 
\usepackage{amsmath}
\usepackage{bm}            
\usepackage{hyperref} 

\usepackage{amsmath}
\usepackage{graphicx}

\makeatletter
\usepackage{mathtools}
\usepackage{cuted}

\usepackage[british]{babel}
\begin{document}

\title{Heat capacity of anisotropic Heisenberg antiferromagnet within the
	spin Hartree-Fock approach in quasi-1D Regime}
\titlerunning{Heat Capacity of Anisotropic Heisenberg Antiferromagnet within the
	Spin Hartree-Fock Approach...}

\author{R. Smit \and P. Kopietz \and O. Tsyplyatyev\thanks{Corresponding author: e-mail o.tsyplyatyev@gmail.com}
}

\institute{Institut f\"ur Theoretische Physik, Universit\"at Frankfurt,
	Max-von-Laue Strasse 1, 60438 Frankfurt, Germany}

\date{Received: date / Revised version: date}


\abstract{We study the anisotropic quantum Heisenberg
antiferromagnet for spin-1/2 that interpolates smoothly between the
one-dimensional (1D) and the two-dimensional (2D) limits. Using the spin Hartree-Fock approach we construct a quantitative theory of heat capacity in the quasi-1D regime with a finite coupling between spin chains. This theory reproduces closely the exact result of Bethe Ansatz in the 1D limit and does not produces any spurious phase transitions for any anisotropy in the quasi-1D regime at finite temperatures in agreement with the Mermin-Wagner theorem.  We  study the static spin-spin correlation function in order to analyse the interplay of lattice geometry and anisotropy in these systems. We compare the square and triangular lattice. For the latter we find that there is a quantum transition point at an intermediate anisotropy of $\sim0.6$. This quantum phase transition establishes that the quasi-1D regime extends upto a particular point in this geometry.  For the square lattice the change from the 1D to 2D occurs smoothly as a function of anisotropy, \emph{i.e.} it is of the crossover type. Comparing the newly developed theory to the available experimental data on the heat capacity  of $\rm{Cs}_2\rm{CuBr}_4$ and $\rm{Cs}_2\rm{CuCl}_4$ we extract the microscopic constants of the exchange interaction that previously could only be measured  using inelastic neutron scattering in high magnetic fields.    
}

\maketitle

\section{Introduction}

The low-dimensional quantum antiferromagnets have attracted
a significant degree of attention  in the last two decades due to synthesis of an increasing number of
anisotropic magnetic insulators with dimensionalities between one
and and two, such as $\textrm{CsNiCl}_{3}$
\cite{trudeau92},$\textrm{Cs}_{2}\textrm{CuCl}_{4}$ \cite{coldea03,sytcheva09}, and
$\textrm{Cs}_{2}\textrm{CuBr}_{4}$ \cite{ono03}. The dimensionality of these materials is larger than strictly one  where the exact Bethe ansatz methods are available \cite{gaudin_book}
but is below three where the classical antiferromagnetic order was
rigorously proven \cite{Dyson78,Kennedy88}, requiring an accurate
description of the quantum effects in the anisotropic 2D anitferromagnets.
Popular methods to deal with this problem are the Takahashi's modified spin-wave theory \cite{takahashi87}
and the mean-field theory based on the Schwinger-boson representation
of spin-waves by Arovas and Auerbach \cite{arovas_auerbach88}. These approaches are quite successful particularly for ferromagnets, for which they produce the correct subleading terms of the free energy in 1D \cite{takahashi87} that were obtained by means of the exact thermodynamic Bethe ansatz \cite{takahashi71,gaudin71}. However, their predictions are not so good for $S=1/2$ antiferromagnets, for which they predict a spurious phase transition at finite temperatures in 1D and 2D, which is explicitly forbidden in these dimensions by the Mermin-Wagner theorem \cite{mermin_wagner66}. Thus, description of thermodynamic quantities, like the heat capacity, in these systems in the quasi-1D regime, where a finite interaction between the spin chains makes the exact Bethe ansatz already inapplicable, at low to intermediate temperatures  requires a different approach. Other proposed methods  include RVB theory \cite{anderson87,liang88}, Wigner-Jordan fermions \cite{wang92}, interpolation based on the high-temperature expansions \cite{bernu01}, and spin-Hartree Fock approach \cite{werth18}.

In this paper, we use the recently proposed spin Hartree-Fock approach  \cite{werth18} in order to construct a theory for the heat capacity of  the anisotropic Heisenberg model for
spin-1/2 on the square and triangular lattices in the quasi-1D regime with
two different exchange energies $J$ and $J'$ along the direction
of the strongest coupling and in the other direction respectively. Within this theory we evaluate the temperature dependence of heat capacity and establish applicability of the quasi-1D regime for different degrees of anisotropy, using the heat capacity itself at low temperatures and the next-neighbour correlation functions. We asses the quality of this theory quantitatively by fitting the obtained temperature dependence of heat capacity to the available experimental data for $\textrm{Cs}_{2}\textrm{CuCl}_{4}$ and $\textrm{Cs}_{2}\textrm{CuBr}_{4}$. The result for the microscopic exchange energies  in the Heisenberg model obtained using these fits matches well the well-known values obtained in neutron scattering experiment available for this materials \cite{coldea02,coldea03}, confirming the validity of our approach.  
\begin{figure*}[h]
	\begin{center}\includegraphics[width=0.9\textwidth]{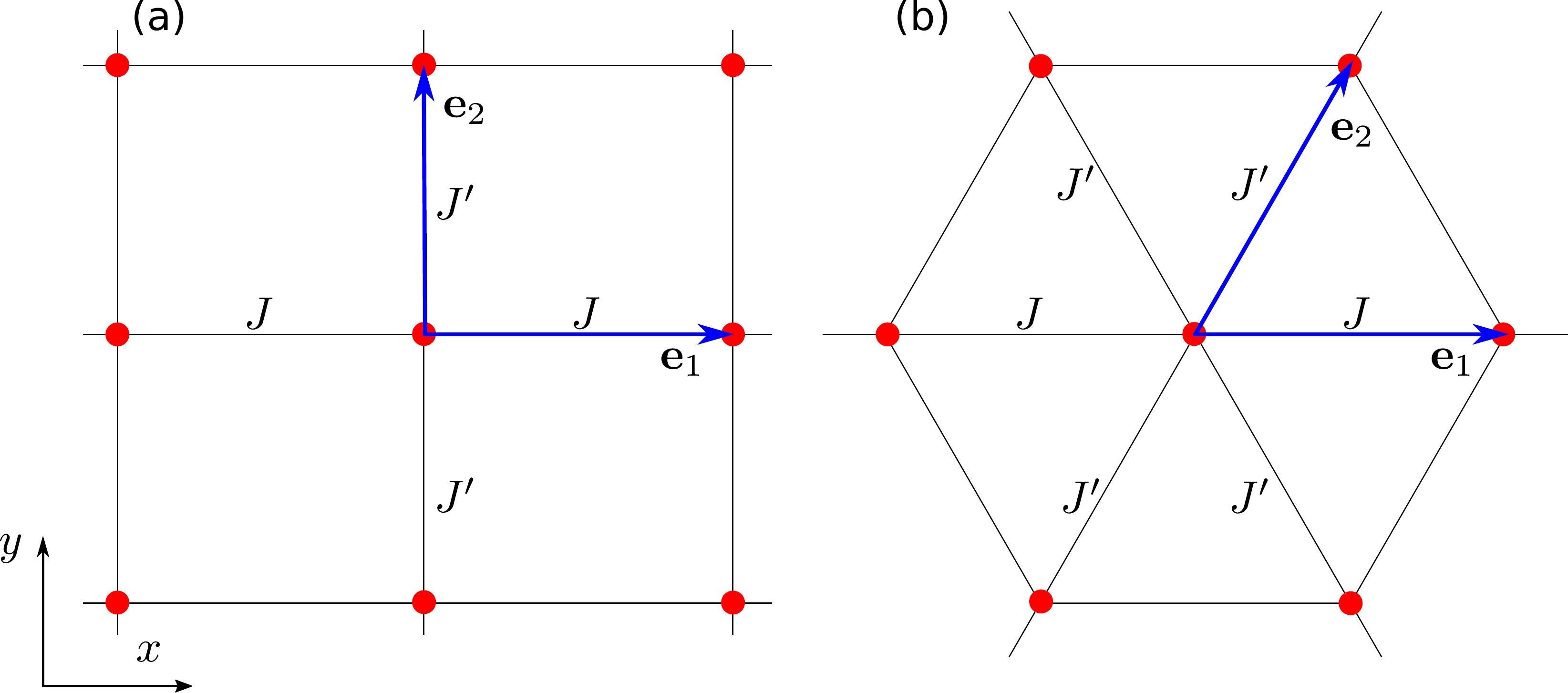}\end{center}\caption{\label{2D_lattices}(a) Anisotropic square lattice formed by spins-1/2
		(red dots). Primitive vectors in the real space are chosen as $\mathbf{e}_{1}=\mathbf{x}$
		and $\mathbf{e}_{2}=\mathbf{y}$. The exchange energy of a given spin-1/2
		with the pair of next-neighbours along the direction of $\mathbf{e}_{1}$
		is $J$ and with the other two next-neighbours is $J'$. (b) Anisotropic
		triangular lattice formed by spins-1/2 (red dots). Primitive vectors
		in the real space are chosen as $\mathbf{e}_{1}=\mathbf{x}$ and $\mathbf{e}_{2}=\mathbf{x}/2+\sqrt{3}\mathbf{y}/2$.
		The exchange energy of a given spin-1/2 with the pair of next-neighbours
		along the direction of $\mathbf{e}_{1}$ is $J$ and with the other
		four next-neighbours is $J'$}
\end{figure*}

The rest of the paper is organised as follows. In Section 2, we describe
the anisotropic Heisenberg model on the square and on the triangular
lattice. Section 3 contains application of the spin Hartree-Fock
approach to the model on the triangular lattice and derivation of
the self-consistency equations for the corresponding mean-field parameters.
In Section 4, we derive the explicit expression for the free energy
within the spin Hartree-Fock approach and analyse the heat capacity
as a function of temperature and anisotropy parameter for the
triangular lattice in detail. In Section 5 we evaluate the static
spin-spin correlation function for the triangular lattice and analyse
in detail its next-neighbour part. In Section 6, we repeat the same
calculations as in Sections 3-5 but for the square lattice and highlight
the differences with the triangular lattice. In Section 6, we fit the heat capacity experimentally measured for a pair magnetic insulators with the theory developed in Sections 3 and 4.

\section{Heisenberg Hamiltonian on anisotropic 2D lattices}

The Heisenberg model for spin-1/2 on the anisotropic 2D lattice in
the absence of an external magnetic field reads 
\begin{equation}
H=\frac{J}{2}\sum_{\mathbf{r},\delta_{1}}\mathbf{S}_{\mathbf{r}}\cdot\mathbf{S}_{\mathbf{r}+\delta_{1}}+\frac{J'}{2}\sum_{\mathbf{r},\delta_{2}}\mathbf{S}_{\mathbf{r}}\cdot\mathbf{S}_{\mathbf{r}+\delta_{2}},\label{eq:H}
\end{equation}
where $J$ and $J'$ are two exchange energies that are different
along a pair of primitive vectors of a 2D lattice, $S_{\mathbf{r}}^{z},S_{\mathbf{r}}^{\pm}=S_{\mathbf{r}}^{x}\pm iS_{\mathbf{r}}^{y}$
are the spin-1/2 operators at site $\mathbf{r}$, the sum over $\mathbf{r}$
runs over a 2D lattice consisting of $N=L^{D}$ spins,
and the sums over $\delta_{1}$ and $\delta_{2}$ run over the nearest-neighbours
connected by the exchange interactions $J$ and $J'$
respectively. We impose the periodic boundary conditions, $\mathbf{S}_{\mathbf{r}+\mathbf{e}_{1}\left(\mathbf{e}_{2}\right)L}=\mathbf{S}_{\mathbf{r}}$,
and restrict ourselves to the antiferromagnetic exchange energies,
$J>0$ and $J'>0$.

In order to compare the effect of different lattice geometries we
consider two particular 2D Bravais lattices in this work. One is
the square lattice as the simplest. The primitive vectors in the units
of lattice parameter are $\mathbf{e}_{1}=\mathbf{x}$ and $\mathbf{e}_{2}=\mathbf{y}$,
see Fig.~\ref{2D_lattices}(a). There are two nearest-neighbours
along the principle direction with the exchange coupling constant
$J$ and two nearest-neighbours in the perpendicular direction with
$J'$. Apart from the square lattice we will also consider the triangular lattice in this work. The primitive vectors in the
units of lattice parameter are $\mathbf{e}_{1}=\mathbf{x}$ and $\mathbf{e}_{2}=\mathbf{x}/2+\sqrt{3}\mathbf{y}/2$,
see Fig.~\ref{2D_lattices}(b). There are two nearest-neighbours
along the principle direction with the exchange coupling constant
$J$ and the other four nearest-neighbours out of the total of six
on this lattice with $J'$. The precise phase diagram for arbitrary $J'/J$ is still not fully established, see a review in \cite{balents10}. Below, we consider the triangular lattice
in detail motivated by a pair of magnetic insulators, $\textrm{Cs}_{2}\textrm{Cu}\textrm{Cl}_{4}$
 and $\textrm{Cs}_{2}\textrm{Cu}\textrm{Br}_{4}$.  The analysis of the square lattice is very
similar and is concisely presented in Section 6. 

\section{Spin Hartree-Fock approach}
In order to analyse the model in Eq.~(\ref{eq:H}) we employ the
spin Hartree-Fock approach developed in Ref. \cite{werth18}. Using
this method in the present Section we derive the self-consistency equations
for the mean-field parameters in the anisotropic case for the triangular
lattice. Solutions of these equations as a function of temperature
and the degree of anisotropy are used later to analyse the thermodynamics
and correlation functions in Sections 5 and 6.

The number of the independent spin components in Eq.~(\ref{eq:H})
can be reduced since the spin-1/2 operators on the same site satisfy
the following identity, 
\begin{equation}
S_{\mathbf{r}}^{z}=S_{\mathbf{r}}^{+}S_{\mathbf{r}}^{-}-1/2,\label{eq:sz1/2}
\end{equation}
in addition to the angular momentum commutation relations $\left[S_{\mathbf{r}}^{\alpha},S_{\mathbf{r}}^{\beta}\right]=i\varepsilon_{\alpha\beta\gamma}S_{\mathbf{r}}^{\gamma}$,
where $\varepsilon_{\alpha\beta\gamma}$ is the Levi-Civita symbol
of third rank. Substituting Eq.~(\ref{eq:sz1/2}) in the $z$-component
of the scalar products in Eq.~(\ref{eq:H}) and performing a Fourier
transform of the whole Hamiltonian by substituting $S_{\mathbf{r}}^{\pm}=N^{-1/2}\sum_{\mathbf{k}}S_{\mathbf{k}}^{\pm}e^{\pm i\mathbf{k}\cdot\mathbf{r}}$
 we obtain a sum of a quadratic and a
quartic form in the spin operators in the momentum space, 
\begin{multline}
H=\sum_{\mathbf{k}}\left[J\left(\varepsilon_{1}\left(\mathbf{k}\right)-1\right)+2J'\left(\varepsilon_{2}\left(\mathbf{k}\right)-1\right)\right]S_{\mathbf{k}}^{+}S_{\mathbf{k}}^{-}\\
+\frac{1}{N}\sum_{\mathbf{k}_{1},\mathbf{k}_{3},\mathbf{k}_{2},\mathbf{k}_{4}}\left[J\varepsilon_{1}\left(\mathbf{k}_{3}-\mathbf{k}_{4}\right)+2J'\varepsilon_{2}\left(\mathbf{k}_{3}-\mathbf{k}_{4}\right)\right]\\
\times\delta_{\mathbf{k}_{1}+\mathbf{k}_{3},\mathbf{k}_{2}+\mathbf{k}_{4}}S_{\mathbf{k}_{1}}^{+}S_{\mathbf{k}_{2}}^{-}S_{\mathbf{k}_{3}}^{+}S_{\mathbf{k}_{4}}^{-},\label{eq:Hk}
\end{multline}
where $\varepsilon_{1}\left(\mathbf{k}\right)=\cos k_{x}$ and $\varepsilon_{2}\left(\mathbf{k}\right)=\cos\left(k_{x}/2\right)$\linebreak $\cos\left(\sqrt{3}k_{y}/2\right)$
are the dispersions along the primitive vectors $\mathbf{e}_{1}$
and $\mathbf{e}_{2}$.  Here we set the lattice and the Planck constants to unity, $a=1$ and $\hbar=1$ respectively.

The structure of the model in Eq.~(\ref{eq:Hk}) resembles that of
a model of interacting particles but for spin operators. The first
(quadratic term) could be interpreted as a kinetic energy and the
second (quartic term) as a two-body interaction between magnetic excitations.
Thus, we analyse the model in Eq.~(\ref{eq:Hk}) using a spin variety of the Hartree-Fock approximation.
Following Ref.~\cite{werth18}, we make an assumption that the many-body
eigenstates can be approximated by product states of the single spin
magnetic excitations when the number of spins $N$ is very large in
the thermodynamic limit. Taking into account finite temperature, such
an approximation corresponds to a product density matrix, 
$\rho=\prod_{\mathbf{k}}\left[m_{\mathbf{k}}\left|\mathbf{k}\rangle\langle\mathbf{k}\right|+1-m_{\mathbf{k}}\right],$
where $\left|\mathbf{k}\right\rangle $ is a state of a single magnetic
excitation with momentum $\mathbf{k}$, the scalar parameters $m_{\mathbf{k}}$
are free parameters that will be defined later, and the normalisation
of the density matrix is chosen such that $\textrm{Tr}\rho=1$. 

The free parameters $m_{\mathbf{k}}$ can obtained in the usual to
mean-field way by minimising the free energy, 
\begin{equation}
F=E-TS,\label{eq:F_def}
\end{equation}
at an arbitrary temperature $T$. The expectation
value of Eq.~(\ref{eq:Hk}) with respect to $\rho$ gives the
expression for the energy of the system $E=\textrm{Tr}\left[H\rho\right]$
as a function of parameters $m_{\mathbf{k}}$, 
\begin{multline}
E=\sum_{\mathbf{k}}\left[J\left(\varepsilon_{1}\left(\mathbf{k}\right)-1\right)+2J'\left(\varepsilon_{2}\left(\mathbf{k}\right)-1\right)\right]m_{\mathbf{k}}\\
-\frac{1}{N}\sum_{\mathbf{k}_{1},\mathbf{k}_{2}}\left(J\varepsilon_{1}\left(\mathbf{k}_{1}-\mathbf{k}_{2}\right)+2J'\varepsilon_{2}\left(\mathbf{k}_{1}-\mathbf{k}_{2}\right)\right)m_{\mathbf{k}_{1}}m_{\mathbf{k}_{2}}.\label{eq:E}
\end{multline}
The von Neumann entropy, $S=-k_{B}\textrm{Tr}\left[\rho\ln\rho\right]$, can be expressed in terms of the parameters $m_{\mathbf{k}}$ as follows,
\begin{equation}
S=-k_{B}\sum_{\mathbf{k}}\left[m_{\mathbf{k}}\ln m_{\mathbf{k}}+\left(1-m_{\mathbf{k}}\right)\ln\left(1-m_{\mathbf{k}}\right)\right],\label{eq:S_def}
\end{equation}
where $k_{B}$ is the Boltzmann constant. Then the solution of the minimisation
problem, $\partial F/\partial m_{\mathbf{k}}=0$, gives  the self-consistency
equations for each $m_{\mathbf{k}}$ in the form of a large set of
$N$ nonlinear equations,
\begin{equation}
m_{\boldsymbol{k}}=\frac{1}{\exp\left[\frac{\mathcal{E}\left(\mathbf{k}\right)}{k_{B}T}\right]+1},\label{eq:mk}
\end{equation}
where the energy in the exponential for every $m_{\mathbf{k}}$ depends
on the all other mean-field parameters as 
\begin{multline}
\mathcal{E}\left(\mathbf{k}\right)=J\left(\varepsilon_{1}\left(\mathbf{k}\right)-1\right)+2J'\left(\varepsilon_{2}\left(\mathbf{k}\right)-1\right)\\
-\frac{2}{N}\sum_{\mathbf{k}'}\left[J\varepsilon_{1}\left(\mathbf{k}-\mathbf{k}'\right)+2J'\varepsilon_{2}\left(\mathbf{k}-\mathbf{k}'\right)\right]m_{\mathbf{k}'}.\label{eq:E1}
\end{multline}

Since $m_{\mathbf{k}}$ in the set of equations in Eq. (\ref{eq:mk}) enter
only under a sum, the number of equations can be reduced to just a
few. Introduction of the following three extensive variables, 
\begin{align}
s= & \frac{1}{N}\sum_{\mathbf{k}}m_{\mathbf{k}}-\frac{1}{2},\label{eq:s_def}\\
u_{\alpha}=- & \frac{1}{N}\sum_{\mathbf{k}}\varepsilon_{\alpha}\left(\mathbf{k}\right)m_{\mathbf{k}}+\frac{1}{2},\label{eq:u_def}
\end{align}
\pagebreak \begin{strip}where $\alpha=1,2$ corresponds to the primitive vectors $\mathbf{e}_{1}$
and $\mathbf{e}_{2}$, reduces Eq.~(\ref{eq:mk}) to a set of only
three independent equations,
\begin{align}
s & =\int\frac{d^{2}k}{V}\frac{1}{\exp\left[\left(\left(2J+4J'\right)s+2Ju_{1}\varepsilon_{1}\left(\mathbf{k}\right)+4J'u_{2}\varepsilon_{2}\left(\mathbf{k}\right)\right)/\left(k_{B}T\right)\right]+1}-\frac{1}{2},\label{eq:s}\\
u_{\alpha} & =\frac{1}{2}-\int\frac{d^{2}k}{V}\frac{\varepsilon_{\alpha}\left(\mathbf{k}\right)}{\exp\left[\left(\left(2J+4J'\right)s+2Ju_{1}\varepsilon_{1}\left(\mathbf{k}\right)+4J'u_{2}\varepsilon_{2}\left(\mathbf{k}\right)\right)/\left(k_{B}T\right)\right]+1},\label{eq:ualpha}
\end{align}
\end{strip}where the thermodynamic limit was taken as $\sum_{\mathbf{k}}/N\rightarrow\int d^{2}k/V$,
the momentum integral runs over the first Brillouin zone, and $V=8\pi^{2}/\sqrt{3}$
is the volume of the primitive cell in the reciprocal space of the
triangular lattice. A renormalised single-particle like dispersion
in the exponent of the equations in Eq.~(\ref{eq:s}) and (\ref{eq:ualpha})
also reproduces qualitatively the Bethe ansatz result \cite{tsyplyatyev14}
and gives the leading contribution to observables in the thermodynamic
limit at high energy \cite{tsyplyatyev15,tsyplyatyev16,moreno16}.

The system of nonlinear equations in Eq.~(\ref{eq:s}) and (\ref{eq:ualpha})
can be solved numerically. For $J'=0$ the equation for $u_{2}$ drops
out (the value $u_{2}\equiv1/2$ becomes independent of all other
parameters) since the system becomes one-dimensional. The two remaining
equations for $s$ and $u_{1}$ can be solved explicitly at zero temperature,
$T=0.$ The integrands are proportional to the Heaviside step function,
\begin{equation}
\lim_{T\rightarrow0}\frac{1}{e^{x/\left(k_{B}T\right)}+1}=\Theta\left(-x\right),
\end{equation}
and the integrals can be evaluated explicitly. Solution of the resulting
linear equations immediately gives only one solution $s=0$ and $u_{1}=1/2+1/\pi$.
Starting from this point Eqs.~(\ref{eq:s}) and (\ref{eq:ualpha})
can be continuously deformed in a smooth way to arbitrary values of
$J'/J$ and $T$ giving $s$ and $u_{\alpha}$ as functions of $J'/J$
and $T$. For all values of $J'/J$ and $T$ this procedure gives
$s=0$, \emph{i.e.} the net magnetisation of the antiferromagnetic
system is always zero. The thermodynamic observables and the static
correlation functions can be expressed in terms of the two functions
$u_{1,2}\left(J'/J,T\right)$. We will analyse some representative
examples of them in the next two Sections.

\section{Thermodynamics}

\begin{figure*}
\begin{center}\includegraphics[width=1\textwidth]{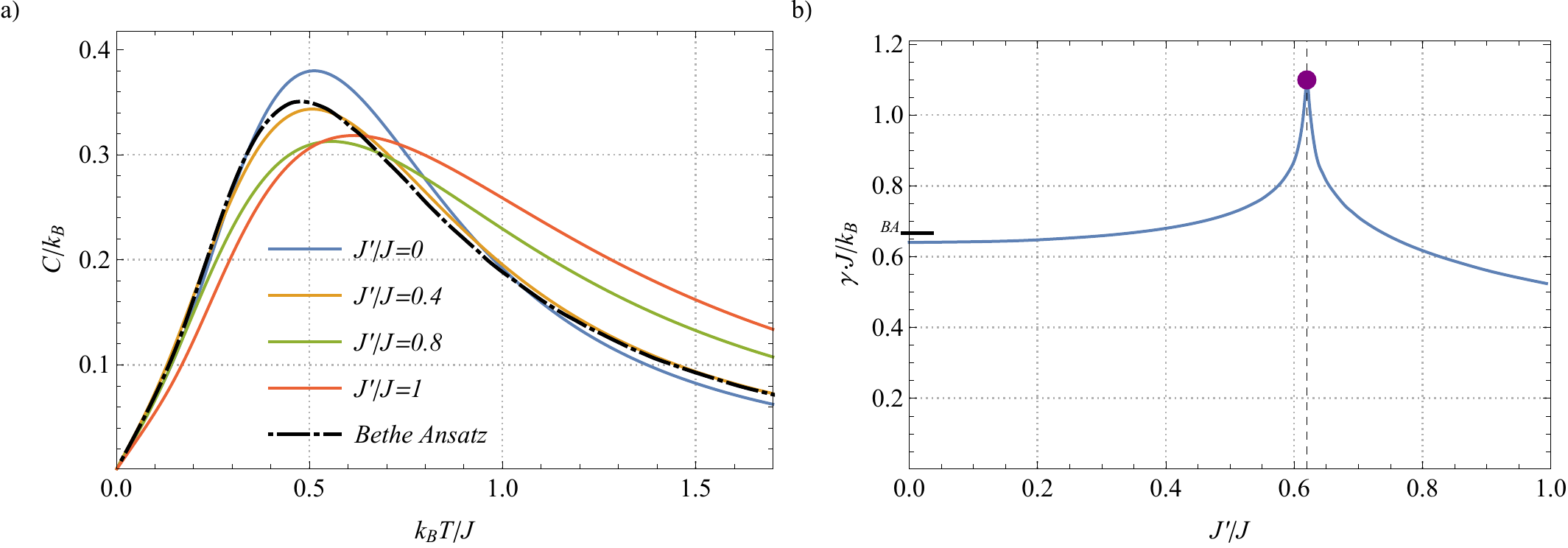}\end{center}\caption{\label{heat_capacity_triangular}(a) Specific heat $C$ as a function of $T$ for triangular lattice for a set of different values of the anisotropy parameter: $J'/J=0$ (blue line), $0.4$ (yellow), $0.8$ (green), and $1$ (red). The curves are obtained numerically by solving the self-consistency equations in Eq.~(\ref{eq:s}) and (\ref{eq:ualpha}) and using the expression for the heat capacity in Eq.~(\ref{eq:C}). The dash-dotted line is the result of Bethe ansatz from \cite{klumper93}. (b) The dependence of the linear coefficient in the heat capacity at low temperatures, $C=\gamma T$, as a function of the anisotropy parameter $J'/J$ obtained numerically using the same approach as in (a). The thick black line at $J'/J=0$ is the result of the Bethe ansatz calculation from \cite{takahashi73}. The black dashed line marks  an intermediate anisotropy at which a cusp occurs. A finite value of $\gamma J/k_B= 1.0991\dots $ at the cusp is marked by a full violet circle.}
\end{figure*}

Now we use the mean-field equations obtained in the previous Section
to analyse the thermodynamic properties of the model in Eq.~(\ref{eq:H}).
First, we derive the general expression for the free energy in terms
of the mean-field parameters $s$ and $u_{\alpha}$ that can be used
to evaluate all thermodynamic quantities. Then, we use a particular
thermodynamic observable, the specific heat $C$, to study in detail
the dependencies on the anisotropy parameter $J'/J$ and temperature $T$. 

The free energy in Eq.~(\ref{eq:F_def}) is expressed in terms of
$E$ and $S$ that, in turn, can be expressed through the three mean-field
parameters $s$ and $u_{\alpha}$ and temperature $T$. Using the
definitions in Eqs.~(\ref{eq:s_def}) and (\ref{eq:u_def}), the
energy of the system $E$ in Eq.~(\ref{eq:E}) can be written as 

\begin{equation}
E=N\left[\left(J+2J'\right)s^{2}-J u_{1}^{2}-2J' u_{2}^{2}\right].\label{eq:E_mf}
\end{equation}
The expression for the entropy is more complicated. The energy in the
exponential in Eq.~(\ref{eq:E1}) can be expressed, analogously to
Eq.~(\ref{eq:E_mf}), using the definitions in Eqs.~(\ref{eq:s_def})
and (\ref{eq:u_def}) in a simple way as 
\begin{equation}
\mathcal{E}\left(\mathbf{k}\right)=\left(2J+4J'\right)s+2Ju_{1}\varepsilon_{1}\left(\mathbf{k}\right)+4J'u_{2}\varepsilon_{2}\left(\mathbf{k}\right).\label{eq:E1_mf}
\end{equation}
The von Neumann entropy in Eq. (\ref{eq:S_def}) then can be written
in terms of this $\mathcal{E}\left(\mathbf{k}\right)$ as 
\begin{multline}
S=Nk_{B}T\int\frac{d^{2}k}{V}\tanh\left(\frac{\mathcal{E}\left(\mathbf{k}\right)}{2k_{B}T}\right)\mathcal{E}\left(\mathbf{k}\right)\\
+Nk_{B}T\int\frac{d^{2}k}{V}\ln\left[\cosh\left(\frac{\mathcal{E}\left(\mathbf{k}\right)}{2k_{B}T}\right)\right].\label{eq:S_mf}
\end{multline}
Substitution of Eqs.~(\ref{eq:E_mf}-\ref{eq:S_mf}) into Eq.~(\ref{eq:F_def})
gives the expression for the free energy in terms of only the mean-field
parameters, the microscopic parameters of the original model in Eq.~(\ref{eq:H}),
and temperature. All the thermodynamic observables can be derived
from it using the general thermodynamic identities and derivatives
with respect to microscopic parameters and temperature. However, the
resulting expressions obtained using this generic method are not very
compact and are rather complicated due to the integrals over momentum in the expression
for entropy in Eq.~(\ref{eq:S_mf}).

Here we focus on the specific thermodynamic observa\-ble---the heat
capacity $C$. It can be expressed through $s$ and $u_{\alpha}$
in a simpler way using its original definition, 
\begin{equation}
C=\frac{\partial E}{\partial T},\label{eq:C_basic_def}
\end{equation}
instead of taking a second derivative of the free energy obtained above,
\begin{equation}
C=-T\frac{\partial^{2}F}{\partial T^{2}}.
\end{equation}
Substitution of the energy $E$ in Eq.~(\ref{eq:E_mf}) into Eq.~(\ref{eq:C_basic_def})
gives the following expression for the heat capacity per spin,
\begin{equation}
\frac{C}{N}=\left(2J+4J'\right)\frac{\partial s}{\partial T}-2Ju_{1}\frac{\partial u_1}{\partial T}-4J'u_{2}\frac{\partial u_{2}}{\partial T}.\label{eq:C}
\end{equation}
This expression, together with the numerical solutions of Eqs.~(\ref{eq:s})
and (\ref{eq:ualpha}) for various values of $T$ and $J'/J$ and where the derivative with respect to temperature is evaluated numerically, is
used for obtaining the plots in Fig. \ref{heat_capacity_triangular}.

The temperature dependence of $C$ for different values of the anisotropy
$J'/J$ ranging from $J'/J=0$ (the 1D limit) to $J'/J=1$ (the
2D limit) is plotted in Fig. \ref{heat_capacity_triangular}(a).
In the 1D limit a 2D magnet splits into a set of independent
1D magnets that are completely isolated from each other. For each
1D magnet the 1D Heisenberg model for spin-1/2 can be diagonalised
exactly using Bethe ansatz \cite{bethe31} and the heat capacity can
also be evaluated using this diagonalisation procedure without any
approximation \cite{takahashi73,klumper93}. The exact result reproduces
quite well the $J'/J=0$ curve in Fig. \ref{heat_capacity_triangular}(a),
especially at low temperatures up to $T/J\sim0.35$. In this region
quantitative difference between the exact result and the result of the spin Hartee-Fock
approach is negligible. In the region from $T/J\approx 0.35$ to $T/J\approx 1$ deviations
are still appreciable, of the order of 15\%. The accuracy in this region can be improved by taking into account systematically the correlation function of order higher than two, {\emph e.g.} using the recently proposed spin-FRG approach \cite{Kopietz18}. 

For all values of the anisotropy  there is no sign
of any finite temperature phase transition, \emph{i.e.} there are
no signs of a singularity or of a kink. This behaviour is in full
accord with the Mermin-Wagner theorem \cite{mermin_wagner66} that
explicitly forbids a long-range order parameter for the spin-1/2 Heisenberg
magnets at finite temperatures in two and lower dimensions. In the low temperature limit, $T\rightarrow0$, the heat capacity remains
a linear function of $T$, $C=\gamma T$, when a small but finite $J'$ is introduced.
The dependence of $\gamma$ on
the anisotropy parameter $J'/J$ is plotted in Fig. \ref{heat_capacity_triangular}(b). At an intermediate anisotropy of $J'\approx 0.6198 J$ the linear coefficient at low temperatures has a cusp signalling a quantum phase transition for the triangular lattice, which establishes the validity of the quasi-1D regime. It will be analysed in more detail using the static correlation function that we evaluate in the next section. 

The conventional spin-wave theory gives a quadratic in $T$ dependence in the 2D regime for $J'>0.6198 J$. However, we are concentrated on the quasi-1D regime $J<0.6198 J$ in this work. This value of $J'_{\mathrm{crit}}=0.6198 J$ is close to values of $J'_{\mathrm{crit}}\approx 0.6 J$ obtained by other methods \cite{yunoki06,hayashi07,kohno07}.

\section{Correlation functions}

In this Section we analyse the static spin-spin correlation function of the
model in Eq.~(\ref{eq:H}). We derive the general expression for
the static finite-range spin-spin correlation function within the
Hartree-Fock approach. Then, we study in detail its next-neighbour
part along the principal axis of the anisotropic triangular
lattice as a function of the anisotropy parameter $J'/J$.

The static spin-spin correlation function can be evaluated as an expectation
value of the spin operator $\mathbf{S}_{0}\cdot\mathbf{S}_{\mathbf{r}}$
with respect to the density matrix in Eq.~(\ref{eq:mk}). The result
depends on many mean-field parameters $m_{\mathbf{k}}$. Using the
definitions in Eqs.~(\ref{eq:s}) and (\ref{eq:ualpha}), the expression
can be reduced to only a function of the three extensive parameters
$s$ and $u_{\alpha}$,
\begin{equation}
\left\langle \mathbf{S}_{0}\cdot\mathbf{S}_{\mathbf{r}}\right\rangle =s^{2}+I\left(\mathbf{r}\right)\left[1-I\left(\mathbf{r}\right)\right],\label{eq:cf_triangular}
\end{equation}
where the function that depends on the coordinate is\begin{strip}
\begin{equation}
I\left(\mathbf{r}\right)=\int\frac{d^{2}k}{V}\frac{\cos\left(\mathbf{k}\cdot\mathbf{r}\right)}{\exp\left[\left(\left(2J+4J'\right)s+2Ju_{1}\varepsilon_{1}\left(\mathbf{k}\right)+4J'u_{2}\varepsilon_{2}\left(\mathbf{k}\right)\right)/\left(k_{B}T\right)\right]+1}\label{eq:Ir_triangular}
\end{equation}
\end{strip}Note that the coordinate $\mathbf{r}$ above labels a
discrete set of nodes on the triangular lattice, i.e. $\mathbf{r}=\mathbf{e}_{1}i+\mathbf{e}_{2}j$
where $i$ and $j$ integer numbers. At $T=0$ the expression in Eqs.~(\ref{eq:cf_triangular})
and (\ref{eq:Ir_triangular}) gives correlations that vanish as a
power-law and at finite $T$ it gives produces an exponential behaviour
at long distances, see details and discussion in Ref. \cite{werth18}.

The signatures of the classical order, however, still survive in the
next-neighbour part of the correlation function. In this case the
integral in Eq.~(\ref{eq:Ir_triangular}) simplifies even further
by use of the definitions in Eqs.~(\ref{eq:s_def}) and (\ref{eq:u_def})
giving $I\left(\mathbf{e}_{1}\right)=1/2-u_{1}$ and $I\left(\mathbf{e}_{2}\right)=1/2-u_{2}$.
Then, the next-neighbour correlation function along the principle
direction of the triangular lattice in Fig.~\ref{2D_lattices}(b)
obtained by the substitution of these expressions into Eq. (\ref{eq:cf_triangular})
reads as 
\begin{equation}
\left\langle \mathbf{S}_{0}\cdot\mathbf{S}_{\mathbf{e}_{1}\left(\mathbf{e}_{2}\right)}\right\rangle =s^{2}-u_{1\left(2\right)}^{2}+\frac{1}{4}.\label{eq:corr_func_nn}
\end{equation}

\begin{figure}[t]
	\begin{center}\includegraphics[width=1\columnwidth]{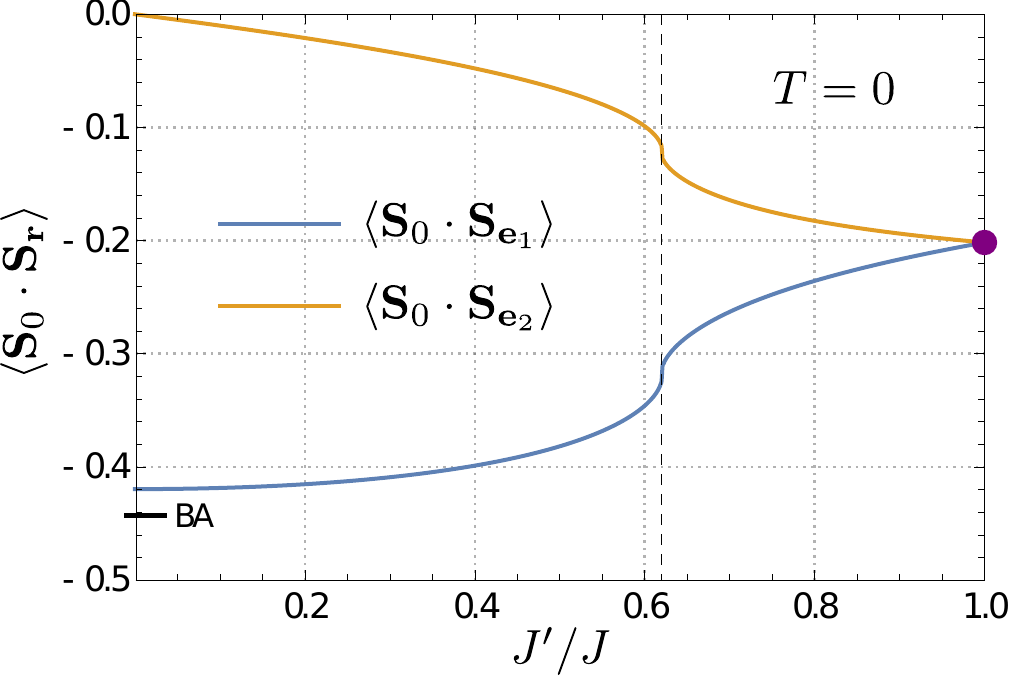}\end{center}\caption{\label{corr_func_triangular}The next-neighbour part of the static spin-spin correlation
		function for the anisotropic triangular lattice as a function of the
		anisotropy parameter $J'/J$ obtained numerically by solving the self-consistency
		equations in Eq.~(\ref{eq:s}) and (\ref{eq:ualpha}) at $T=0$ and, then,
		by using the expression in Eq.~(\ref{eq:corr_func_nn}). The blue
		solid line is the correlation function along the chains $\left\langle \mathbf{S}_{0}\cdot\mathbf{S}_{\mathbf{e}_{1}}\right\rangle $
		and the yellow solid line is the correlation function in the other principle direction $\left\langle \mathbf{S}_{0}\cdot\mathbf{S}_{\mathbf{e}_{1}}\right\rangle $.
		The black line is the exact value obtained by Bethe ansatz for 
		the 1D system and the black dashed line is the quantum transition
		point $J'\approx 0.6198 J$ separating the quasi-1D from the 2D order. The full violet circle is  $\left\langle \mathbf{S}_{0}\cdot\mathbf{S}_{\mathbf{e}_{1}}\right\rangle =\left\langle \mathbf{S}_{0}\cdot\mathbf{S}_{\mathbf{e}_{2}}\right\rangle =-0.2017\dots$ for $J'/J=1$.}
\end{figure}

The above expression with the solution of Eqs.~(\ref{eq:s}) and
(\ref{eq:ualpha}) as a function of the anisotropy $J'/J$ at zero
temperature $T=0$ is plotted in Fig.~\ref{corr_func_triangular}.
In the 1D limit ($J'/J=0$) the next-neighbour correlation function
along the chains (blue line) is $\left\langle \mathbf{S}_{0}\cdot\mathbf{S}_{\mathbf{e}_{1}}\right\rangle =-0.4196\dots$
that is close to the Bethe ansatz result $\left\langle \mathbf{S}_{0}\cdot\mathbf{S}_{\mathbf{e}_{1}}\right\rangle =-0.4431\dots$
\cite{orbach58} marked by the thick black line. 
\begin{figure}[b]
\begin{center}\includegraphics[width=1\columnwidth]{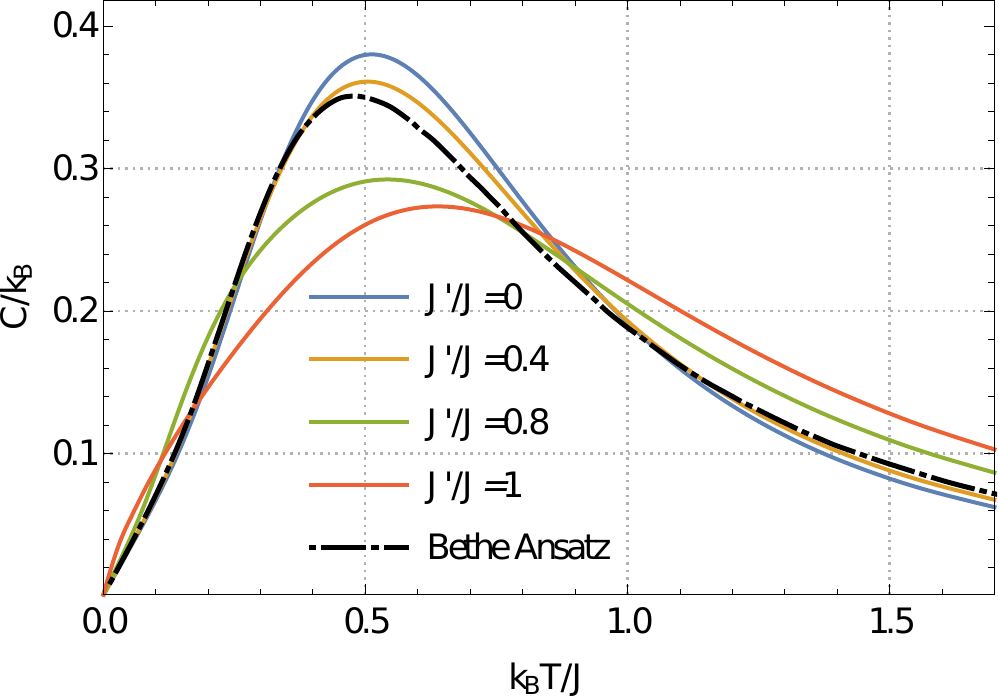}\end{center}\caption{\label{heat_capacity_square} Specific heat $C$ as a function
of $T$ for the anisotropic square lattice for a set of different values of the anisotropy parameter
$J'/J=0$ (blue line), $0.4$ (yellow line), $0.8$ (green line), and $1$ (red line). The curves are
obtained numerically by solving the self-consistency equations in Eq.~(\ref{eq:s_square})
and (\ref{eq:ualpha_square}) and using the expression for the heat
capacity in Eq.~(\ref{eq:C_square}). The dash-dotted line is the result of Bethe ansatz from \cite{klumper93}. }
\end{figure}

In the $J'/J=1$ limit  the next-neighbour correlation function
in all direction is $\left\langle \mathbf{S}_{0}\cdot\mathbf{S}_{\mathbf{e}_{1}}\right\rangle =\left\langle \mathbf{S}_{0}\cdot\mathbf{S}_{\mathbf{e}_{2}}\right\rangle =-0.2017\dots$
that is reminiscent of the classical 120-degree order $\left\langle \mathbf{S}_{0}\cdot\mathbf{S}_{\mathbf{e}_{1}}\right\rangle =\left\langle \mathbf{S}_{0}\cdot\mathbf{S}_{\mathbf{e}_{2}}\right\rangle =-0.125$
but still shows a relatively large enhancement due to quantum fluctuations. This value is also close the exact diagonalisation result $\left\langle \mathbf{S}_{0}\cdot\mathbf{S}_{\mathbf{e}_{1}}\right\rangle=-0.182\dots$ obtained in \cite{bernu94}.
In between there is a quantum transition point identified as a finite
jump in the correlation function at $J'_{\mathrm{crit}}\approx 0.6198 J$,
which is within the range of $J_{\mathrm{crit}}'$ from $0.60J$ to $0.65J$
obtained by a quantum Monte Carlo approach \cite{yunoki06} and by
using the RVB ansatz \cite{hayashi07,heidarian09}.
Below this point ($J<J_{\textrm{crit}}'$) there is a quasi-1D order
close to the strictly 1D Bethe ansatz solution with weak correlations
between the chains.

\section{Square lattice}

Here we consider the model in Eq.~(\ref{eq:H}) for the anisotropic
square lattice in Fig.~\ref{2D_lattices}(a). We quickly sketch the
application of the spin Hartree-Fock approach in this case that is
almost identical to the triangular lattice considered in Sections
3-5 and quote the corresponding results for the heat capacity and
for the next-neighbour part of the static spin-spin correlation function. Then,
we conclude by analysing numerically these two observables and highlight
the differences in them between the square and the triangular lattices.

In this section we consider the anisotropic square lattice in Fig.~\ref{2D_lattices}(a).
The two primitive vectors in the real space are now orthogonal, $\mathbf{e}_{1}=\mathbf{x}$
and $\mathbf{e}_{2}=\mathbf{y}$. The Fourier transform of the model
in Eq.~(\ref{eq:H}) on this lattice gives
\begin{multline}
H=\sum_{\mathbf{k}}\left[J\left(\varepsilon_{1}\left(\mathbf{k}\right)-1\right)+J'\left(\varepsilon_{2}\left(\mathbf{k}\right)-1\right)\right]S_{\mathbf{k}}^{+}S_{\mathbf{k}}^{-}\\
+\frac{1}{N}\sum_{\mathbf{k}_{1},\mathbf{k}_{3},\mathbf{k}_{2},\mathbf{k}_{4}}\left[J\varepsilon_{1}\left(\mathbf{k}_{3}-\mathbf{k}_{4}\right)+J'\varepsilon_{2}\left(\mathbf{k}_{3}-\mathbf{k}_{4}\right)\right]\\
\times\delta_{\mathbf{k}_{1}+\mathbf{k}_{3},\mathbf{k}_{2}+\mathbf{k}_{4}}S_{\mathbf{k}_{1}}^{+}S_{\mathbf{k}_{2}}^{-}S_{\mathbf{k}_{3}}^{+}S_{\mathbf{k}_{4}}^{-},\label{eq:Hk_square}
\end{multline}
where $\varepsilon_{1}\left(\mathbf{k}\right)=\cos k_{x}$ and $\varepsilon_{2}\left(\mathbf{k}\right)=\cos k_{y}$
are the two dispersions along the primitive vectors $\mathbf{x}$ and
$\mathbf{y}$. Following the Hartree-Fock approach we evaluate the
energy of the system as an expectation value of the model in Eq.~(\ref{eq:Hk_square})
with respect to a thermal state and
obtain the self-consistence equations for the mean-field parameters
$m_{\mathbf{k}}$ by minimising the corresponding free energy. Using
the definitions of the extensive parameters in Eqs.~(\ref{eq:s_def})
and (\ref{eq:u_def}), the set of many ($N$) self-consistency equations
is reduced to only three, similarly to Eqs.~(\ref{eq:s}) and (\ref{eq:ualpha}),\begin{strip}
\begin{align}
s & =\int\frac{d^{2}k}{V}\frac{1}{\exp\left[\left(2\left(J+J'\right)s+2Ju_{x}\varepsilon_{1}\left(\mathbf{k}\right)+2J'u_{y}\varepsilon_{2}\left(\mathbf{k}\right)\right)/\left(k_{B}T\right)\right]+1}-\frac{1}{2},\label{eq:s_square}\\
u_{\alpha} & =\frac{1}{2}-\int\frac{d^{2}k}{V}\frac{\varepsilon_{\alpha}\left(\mathbf{k}\right)}{\exp\left[\left(2\left(J+J'\right)s+2Ju_{x}\varepsilon_{1}\left(\mathbf{k}\right)+2J'u_{y}\varepsilon_{2}\left(\mathbf{k}\right)\right)/\left(k_{B}T\right)\right]+1},\label{eq:ualpha_square}
\end{align}
\end{strip}where $u_{1}\equiv u_{x}$ and $u_{2}\equiv u_{y}$ are
now the variables in the orthogonal directions of the primitive vectors
of the square lattice and $V=\left(2\pi\right)^{2}$ is the volume
of the primitive cell in the reciprocal space of the square lattice.

Analogously to Eq.~(\ref{eq:C}) in Section 4, the heat capacity
per spin can be obtained as a derivative of the energy of the system,
see Eq.~(\ref{eq:C_basic_def}), giving 
\begin{equation}
\frac{C}{N}=2\left(J+J'\right)\partial_{T}s-2Ju_{x}\partial_{T}u_{y}-2J'u_{y}\partial_{T}u_{y}.\label{eq:C_square}
\end{equation}
for the anisotropic square lattice. This expression with the solutions
of Eqs.~(\ref{eq:s_square}) and (\ref{eq:ualpha_square}) is plotted
as a function of temperature $T$ for a set of different anisotropies
in Fig.~\ref{heat_capacity_square}.
\begin{figure}[t]
\begin{center}\includegraphics[width=1\columnwidth]{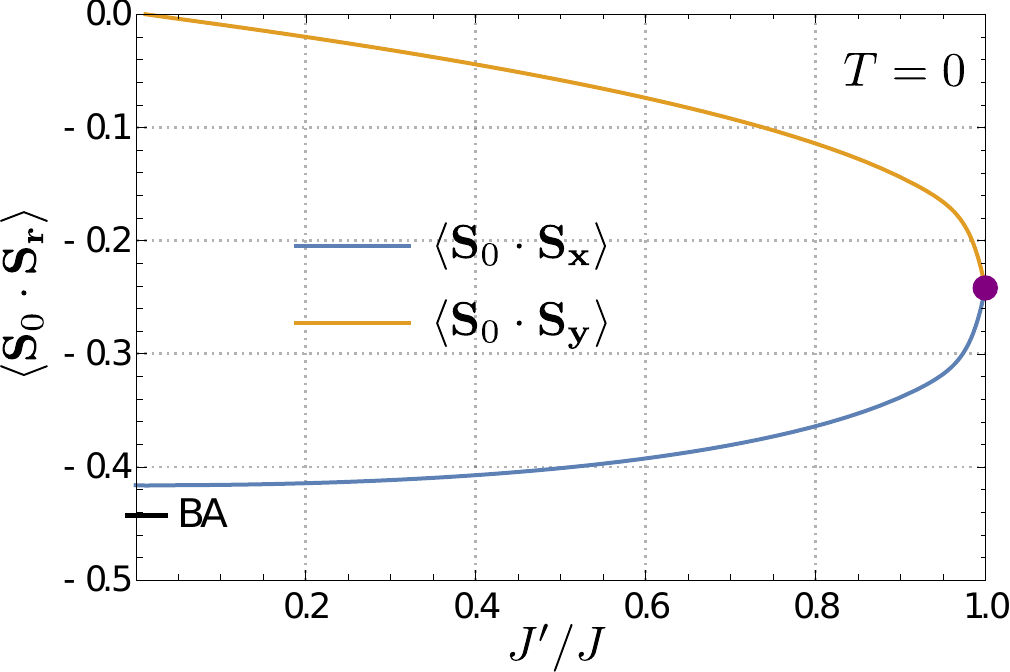}\end{center}\caption{\label{corr_func_square}The next-neighbour part of the static spin-spin correlation
function for the anisotropic square lattice as a function of the anisotropy
parameter $J'/J$ obtained numerically by solving the self-consistency
equations Eq.~(\ref{eq:s_square}) and (\ref{eq:ualpha_square}) at $T=0$
and, then, by using the expression in Eq.~(\ref{eq:corr_func_nn_square}).
The yellow line is the correlation function along the chains $\left\langle \mathbf{S}_{0}\cdot\mathbf{S}_{\mathbf{x}}\right\rangle $
and the blue line is the correlation function in the perpendicular direction $\left\langle \mathbf{S}_{0}\cdot\mathbf{S}_{\mathbf{y}}\right\rangle $.
The thick black line is the exact value obtained by Bethe ansatz for
the 1D system. The full violet circle is  $\left\langle \mathbf{S}_{0}\cdot\mathbf{S}_{\mathbf{x}}\right\rangle =\left\langle \mathbf{S}_{0}\cdot\mathbf{S}_{\mathbf{y}}\right\rangle =-0.2431\dots$ for $J'/J=1$.}
\end{figure}

The next-neighbour part of the static spin-spin correlation function is obtained
in the same way in Section 5 giving the same expression as in Eq.~(\ref{eq:corr_func_nn}),
\begin{equation}
\left\langle \mathbf{S}_{0}\cdot\mathbf{S}_{\mathbf{x}\left(\mathbf{y}\right)}\right\rangle =s^{2}-u_{x\left(y\right)}^{2}+\frac{1}{4}.\label{eq:corr_func_nn_square}
\end{equation}
Note that here the directions of the next-neighbour correlation function
are orthogonal, they are along the primitive vectors of the square
lattice unlike in Section 5. The above expression with the solutions
of Eqs.~(\ref{eq:s_square}) and (\ref{eq:ualpha_square}) is plotted
a function of the anisotropy parameter $J'/J$ at $T=0$ in Fig.~\ref{corr_func_square}.
In the 1D limit ($J'/J=0$) it reproduces closely the exact value
obtained using the available Bethe ansatz approach as in Section 5.
In the 2D limit ($J'/J=1$) the next-neighbour correlation
function $\left\langle \mathbf{S}_{0}\cdot\mathbf{S}_{\mathbf{x}}\right\rangle =\left\langle \mathbf{S}_{0}\cdot\mathbf{S}_{\mathbf{y}}\right\rangle =-0.2431\dots$
is reduced from its classical value $-1/4$
due to quantum fluctuations. This is consistent with the Quantum Monte Carlo result $\left\langle \mathbf{S}_{0}\cdot\mathbf{S}_{\mathbf{x}}\right\rangle = -0.133\dots$ obtained in \cite{sandvik97}. Also for the square lattice there is no
quantum phase transition at any intermediate anisotropy.

\section{Fitting experimental data}

\begin{figure}[h]
	\begin{center}\includegraphics[width=1\columnwidth]{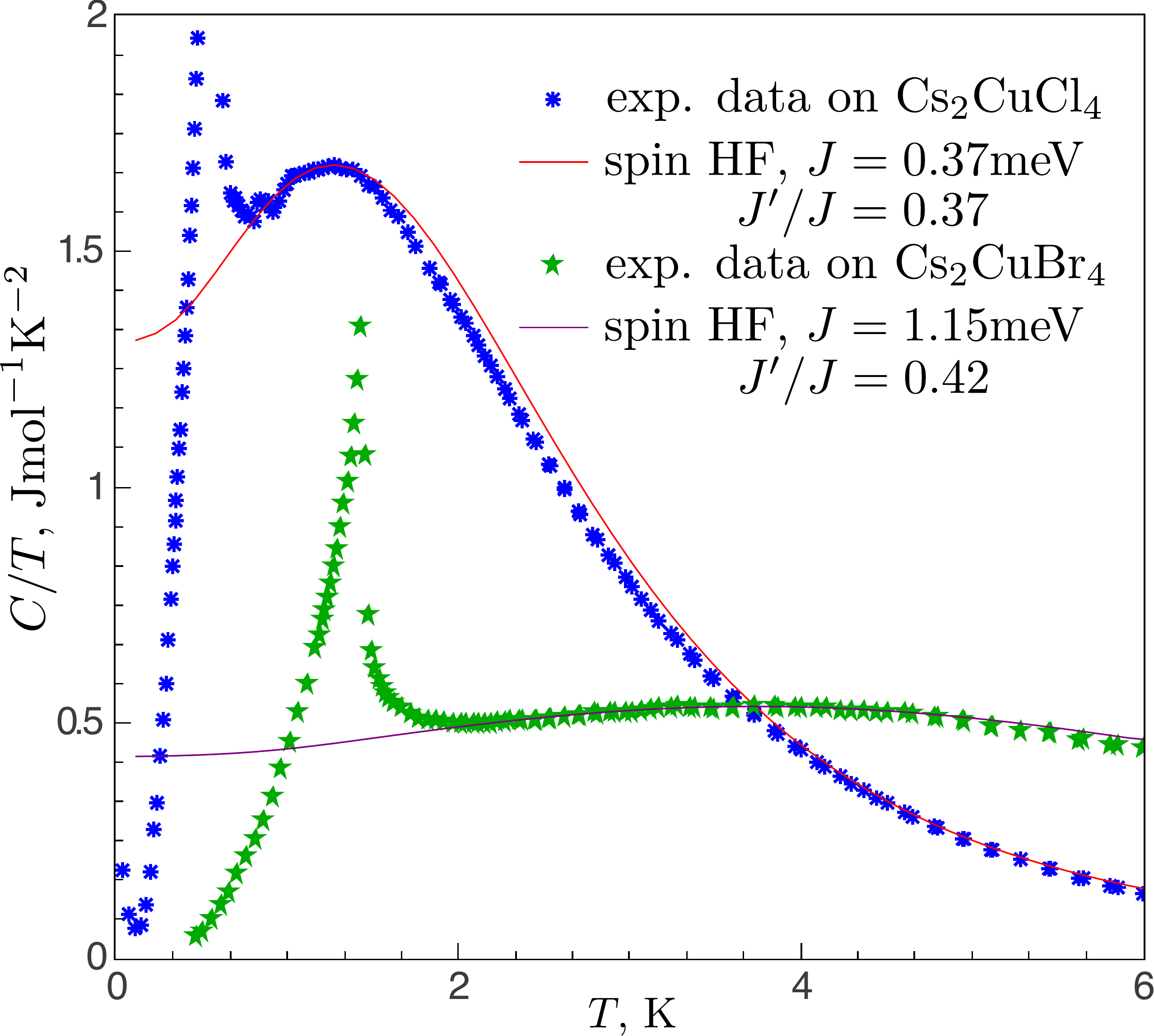}\end{center}\caption{\label{exp_fits} Comparison of the temperature dependence of the heat capacity measured experimentally at zero magnetic field for $\rm{Cs}_2\rm{CuBr}_4$ and $\rm{Cs}_2\rm{CuCl}_4$ in the quasi-1D regime with the theory based on the spin Hartree-Hartree approach developed in this paper. Blue stars are the experimental data for $\rm{Cs}_2\rm{CuCl}_4$ taken from \cite{radu05}. Green stars are the experimental data for $\rm{Cs}_2\rm{CuBr}_4$ taken from \cite{ono03}. Red and purple solid lines are the heat capacity evaluated for anisotropic triangular lattice using Eqs. (\ref{eq:s}), (\ref{eq:ualpha}), and (\ref{eq:C}) for the microscopic parameters $J=0.37$meV, $J'/J=0.37$ and $J=1.15$meV, $J'/J=0.42$ respectively.}  
\end{figure}

Now, we perform a quantitative test of the theory developed in Sections 3 and 4 in the quasi-1D regime by comparing it with the experimental data for heat capacity of the anisotropic quantum antiferromagnets $\rm{Cs}_2\rm{CuBr}_4$ and $\rm{Cs}_2\rm{CuCl}_4$ on the triangular lattice  measured in \cite{ono03} and \cite{radu05} respectively. The experimental data from these two works is plotted in Fig.~\ref{exp_fits}, blue stars are  $\rm{Cs}_2\rm{CuBr}_4$ and green stars are $\rm{Cs}_2\rm{CuCl}_4$. 

Both materials become 3D magnets below a critical temperature that corresponds to a weak Heisenberg coupling between the planes, $T^{3D}_c=2.00$ K for $\rm{Cs}_2\rm{CuBr}_4$ and $T^{3D}_c=0.70$K for $\rm{Cs}_2\rm{CuCl}_4$. Thus, we exclude the data for $T<T^{3D}_c$ from the comparison. We also exclude the data for $T>0.35J/k_B$ since the spin Hartree-Fock approach has a larger quantitative discrepancy in this intermediate region of temperatures, see comparison with the exact result of Bethe ansatz in Fig.~\ref{heat_capacity_triangular}. We also exclude the data at very high temperatures from the fitting procedure since the lattice contribution to the heat capacity in experiments becomes appreciable there making direct comparison of the theory developed in this work for purely magnetic systems inadequate. 

We use the result in Eqs. (\ref{eq:s}), (\ref{eq:ualpha}), and (\ref{eq:C}) to fit the experimental data for both materials in intermediate range of temperatures,  $T^{3D}_c<T<0.35J/k_B$, using $J$ and the degree of anisotropy  $J'/J$ as the only two free fitting parameters, see red and purple lines in Fig.~\ref{exp_fits}. The best fits give   $J=1.15\pm0.01$meV and $J'/J=0.42\pm0.03$ for $\rm{Cs}_2\rm{CuBr}_4$ and  $J=0.37\pm0.01$meV and $J'/J=0.37\pm0.03$ for $\rm{Cs}_2\rm{CuCl}_4$ \cite{other_exps_note}. This procedure was implemented via a minimisation routine using MATHEMATICA, giving also relatively small error bars since the fitting was done well in the nonlinear regime of intermediate temperatures for these materials. The obtained values of microscopic parameters match well the already know results of neutron scattering experiments for these materials at high magnetic fields, where these microscopic parameters of the Heisenberg model are measured directly \cite{coldea02,coldea03}. Thus, accuracy of the theory for heat capacity of quantum antiferromagnets in the quasi-1D regime developed in this work is confirmed.

\section{Conclusions}

We have applied the spin Hartree-Fock approach to the model of the
anisotropic Heisenberg antiferromagnet for spin-1/2 on the square
and  triangular lattices. We have constructed a theory for heat capacity in the quasi-1D regime that is free of any spurious phase transition at finite temperatures, unlike the commonly used  Takahashi's modified spin-wave theory \cite{takahashi87}
and the mean-field theory based on the Schwinger-boson representation
of spin-waves by Arovas and Auerbach \cite{arovas_auerbach88}. We have successfully tested the accuracy of our newly developed approach by fitting the available data on  the temperature dependence of heat capacity  for a pair of anisotropic antiferromagnetic isolators \cite{ono03,radu05} in the quasi-1D regime and obtained the microscopic exchange energies that match the values found in the neutron scattering experiments on these materials, confirming the validity of our theory.

\begin{acknowledgement}
This work was financially supported by the DFG through SFB/TRR 49 and is a contribution to the final report on this project \cite{sfb_report}.
\end{acknowledgement}

\section{Authors contributions}
All  authors  contributed  equally  to  the  paper.


\begin{thebibliography}{[1]}
\bibitem{other_exps_note} The spin Hartree-Fock approached in developed in this work also describes quite well the heat capacity measured in the intermediate compounds $\rm{Cs}_2\rm{CuClBr}_3$ and $\rm{Cs}_2\rm{CuCl}_2\rm{Br}_2$, which were synthesised very recently \cite{tutsch19}.
\end{thebibliography}
\end{document}